# Charge-independent trend of isoscalar matrix elements along the $N \sim Z$ line

J. N. Orce [a] and V. Velázquez [b]

[a] *Department of Physics and Astronomy, University of Kentucky, Lexington, KY 40506-0055, USA*

[b] *Depto. de Física, Facultad de Ciencias, Universidad Nacional Autónoma de México, Apartado Postal 70-543, 04510 México, D.F., México*

**Abstract**

Shell model calculations have been carried out using the m-scheme numerical code ANTOINE in order to elucidate the particular trend of the isoscalar matrix elements, $M_0$, for $A = 4n + 2$ isobaric triplets ranging from $A = 18$ to $A = 42$. The $2^+_{1\,(T=1)} \rightarrow 0^+_{1\,(T=1)}$ transition energies, reduced transition probabilities and isoscalar matrix elements are predicted to a high degree of accuracy. The general agreement of $M_0$ between those from mirror pairs and those from $T_Z = 0$ nuclides support our shell model calculations. The predicted results tie together recent experimental data, and the trend of $M_0$ strength along the $sd$ and beginning of the $fp$ shells is interpreted in terms of the dynamic shell structure. Certain discrepancies arise at $A = 18$ and $A = 38$ isobaric triplets, which might be explained in terms of core polarization effects and the low occupancy of the orbits at the extremes of the $sd$ shell.

## 1 Introduction

That charge independence of the nuclear force in the low energy regime of nuclear structure is an approximate symmetry is confirmed by low energy scattering experiments and similar energies of excited states in isobaric triplets. This

*Email address:* `jnorce@pa.uky.edu` (J. N. Orce).
*URL:* `http://www.pa.uky.edu/~jnorce` (J. N. Orce).

symmetry suggests that $pp$, $np$ and $nn$ forces are equal for the corresponding isobaric states (e.g., $T = 1$ states) and is broken only by the electromagnetic interaction. The less restrictive statement of charge symmetry implies that $nn$ and $pp$ forces are identical and, according to the decay schemes, it is approximately conserved in mirror nuclides up to at least $A \sim 60$ [1]. Assuming that charge symmetry is conserved, the fulfillment of charge independence leads to the conservation of isospin symmetry in nuclear structure [2]. A violation of charge symmetry implies that charge independence is broken but the converse is not true [3].

An ingenious way to examine isospin symmetry, both charge symmetry and charge independence, among isobaric triplets with $A = 4n + 2$, was proposed by Bernstein, Brown and Madsen [4]. In particular, charge independence can experimentally be examined in $T = 1$ isobaric triplets by the confrontation of $2^+_{1\,(T=1)} \rightarrow 0^+_{1\,(T=1)}$ $E2$ transition strengths. In comparison with excitation energies, reduced transition probabilities test nuclear wave functions in greater detail since the latter involve initial and final states. For the same mass number, isospin symmetry could be examined by comparing the isoscalar matrix elements, $M_0$, extracted from the two nuclides with $T_z = \pm 1$ with that extracted from the $T_z = 0$ nuclide. This would test the degree to which the approximate charge independence of nuclear force leads to a single $M_0$ for isobaric triplets. Accordingly, in mirror pairs, $T_z' = -T_z$, the relationship between matrix elements [4] in the neutron or proton isospin representation yields,

$$M_0(T_z) = M_p(T_z) + M_p(-T_z) \tag{1}$$

which implies that for a transition between $T = 1$ states in a $T_z = 0$ nucleus:

$$M_0(T = 1) = 2M_p(T_z = 0) \tag{2}$$

where the proton matrix elements, once the Wigner-Eckart theorem is applied, are given by $M_p = < J_f || \sum r_i^\lambda Y_\lambda(\Omega_i) || J_i >$, and related to the reduced transition probabilities by,

$$M_p = [(2J_i + 1)\ \ B(E2; J_i \rightarrow J_f)]^{1/2}. \tag{3}$$

A recent series of experiments initiated by Cottle *et al.,* and supplemented by others, provide reliable experimental results for $E2$ transition rates in $T = 1$ isobaric triplets through refined techniques of Coulomb excitation and lifetime measurements [5–10]. An earlier apparent problem [5] of broken isospin symmetry for some isobaric triplets between $A = 18$ and $A = 42$ has partially been resolved with certain discrepancies remaining for $A = 34$ and 38. For the case of $A = 18$, this comparison will possibly remain incomplete due to the very short lifetime measured for the first $2^+_{T=1}$ state ($t_{1/2} < 0.83$ fs) in $^{18}$F [11].

In this work we address the intriguing picture of the systematic trend of isoscalar matrix elements for $A = 4n + 2$ isobaric triplets through the *sd* shell, and into the *fp* shell, for nuclei ranging from $A = 18$ to $A = 42$. This systematic trend is examined within a shell model framework. In more detail, from $A = 18$ to $A = 38$, there is a parabolic trend with a maximum for $A = 22$, from which the matrix elements decrease up to $A = 38$, near the $Z = N = 20$ shell closures ($M_0 \sim 8.5$ W.u. for $A = 22$ and $M_0 \sim 2.5$ W.u. for $A = 38$). The trend seems to be broken once the $1f_{7/2}$ shell begins to be filled in the $A = 42$ isobars ($M_0 \sim 7$ W.u.).

In principle, this trend could be tested if one considers that the behaviour of the isoscalar matrix elements is affected by dynamical shell effects. A thorough shell model treatment of isospin symmetry in $E2$ transitions was provided by Brown *et al.*, who examined explicitly the isovector $E2$ transition rates in the *sd* shell [12]. These rates are very sensitive to departures from symmetry, since they correspond to differences in proton matrix elements, $M_p$, of mirror pairs, for example. On the whole, and with the data available at that time, the isovector $E2$ matrix elements followed the trend through the *sd* shell expected from shell model calculations. Examining this trend also enabled extraction of isovector $E2$ effective charges, which account for core polarization effects. Moreover, similar parabolic trends to the isoscalar matrix elements along the *sd* shell have been observed in isotope shifts and $B(E2)$ strengths from $2_1^+$ states to the ground state in the even-$A$ calcium isotopes between shell closures ($^{40}$Ca to $^{48}$Ca), with a maximum for midshell $^{44}$Ca [13]. Through shell model calculations [14], where all configurations of nucleons in the $2s_{1/2}$, $1d_{3/2}$, $1f_{7/2}$ and $2p_{3/2}$ orbits were included in the valence space, it was concluded that the $Z = N = 20$ shell boundaries are not absolute and strong configuration-mixing involving nucleons from the *sd* shell core of $^{40}$Ca into the *fp* shell plays an important role [15]. This has been experimentally confirmed by recent $g$ factor measurements in $^{42}$Ca [16] and $^{44}$Ca [16,17], where positive $g$ values indicate considerable admixtures of core excited configurations in the wave functions. In view of the successful predictions for $^{42}$Ca and the *sd* shell, we have used the $m$-scheme numerical code ANTOINE [14] to elucidate the behaviour of the $A = 4n+2$ isobaric triplets. As noted in ref. [14], this can be applied with relatively large valence spaces.

## 2 Shell model calculations and discussion

In our calculations for the *sd* shell ($\hbar\omega_0$), which include the $1d_{5/2}$, $2s_{1/2}$ and $1d_{3/2}$ subshells in the valence space for protons and neutrons, we used the Wildenthal *USD* interaction [18] and $^{16}$O as the inert core. Calculations without truncations have been performed for the A=38 and A=42 systems using the $2s_{1/2}$, $1d_{3/2}$, $1f_{7/2}$ and $2p_{3/2}$ valence space, with $^{28}$Si as the inert core. From $A=18$ to $A=42$, the calculations were performed with the usual effective charges $e_p=1.5$, $e_n=0.5$, and the harmonic oscillator parameter $b=1.01A^{1/6}$. In order to test primarily the charge-independent trend of experimental $M_0$ values along the N∼Z line, the Coulomb interaction was not considered in our calculations. Figure 1 shows the experimental and calculated energies for the $2_1^+{}_{(T=1)} \rightarrow 0_1^+{}_{(T=1)}$ transitions for mirror pairs and $T_Z$=0 nuclides. As indicated in Brown *et al.*, the $USD$ interaction is an effective choice away from the extremes of the $sd$ shell, presenting good agreements from $A=22$ to $A=34$ (green diamonds). Certain discrepancies arise, however, for the $A=18$ and $A=38$ triplets. These are nicely solved for the $A=38$ system by expanding the valence space into the $fp$ shell and using of the *sdpf.fin* interaction [19], which reproduce the experimental energies for the $A=38$ mirror pair (red square) and substantially improves the calculated $2_1^+{}_{(T=1)} \rightarrow 0_1^+{}_{(T=1)}$ transition energy in $^{38}$K. For the $A=42$ isobaric triplet, we applied a version of the *sdpf.fin* interaction [19] with modified single-particle energies, in agreement with the results of Caurier *et al.* in the light Ca isotopes [15]. Our calculations for $^{42}$Ca are in good agreement with Caurier *et al.* and the $A=42$ $2_1^+{}_{(T=1)} \rightarrow 0_1^+{}_{(T=1)}$ transition energies are also well reproduced. The results for the $A=38$ and $A=42$ isobaric triplets support $sd$ shell excitations into the $fp$ shell and viceversa, as proposed in $^{42}$Ca and $^{44}$Ca. Again, this supports the idea of 'weakened' $N=Z=20$ shell boundaries.

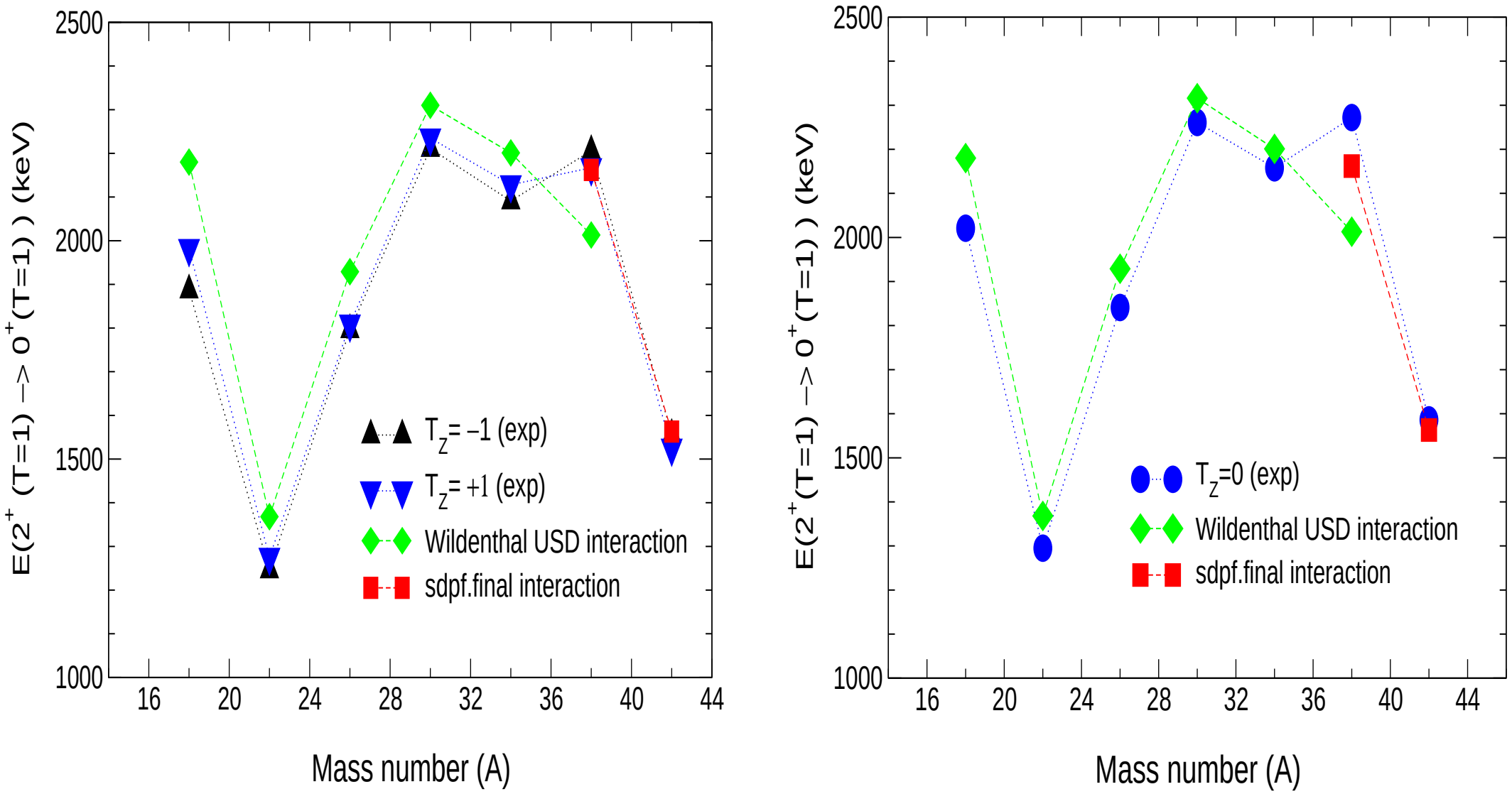

Fig. 1. Experimental and calculated energies for $2_1^+$ $(T=1) \rightarrow 0_1^+$ $(T=1)$ E2 transitions as a function of mass number for $A=4n+2$ (top) mirror nuclei ($T_Z=\pm1$) and (bottom) $T_Z$=0 nuclides along the $N=Z$ line ranging from $A=18$ to $A=42$.

Experimental and calculated $B(E2; 2^+_{1(T=1)} \rightarrow 0^+_{1(T=1)})$ values, proton and isoscalar multipole matrix elements for isobaric triplets ranging from $A=18$ to $A=42$ are listed in Table 1 for comparison. The predicted and experimental

Table 1
Experimental and predicted $B(E2)$ values, proton matrix elements and isoscalar matrix elements for the pure, stretched $2_1^+$ $(T=1) \rightarrow 0_1^+$ $(T=1)$ $E2$ transitions for the $A = 4n+2$ isobaric triplets ranging from $A=18$ to $A=42$ along the line of stability. The contribution to the isoscalar matrix element, $M_0$, from mirror pairs is given by $M_0 = M_p(T_Z=-1) + M_p(T_Z+1)$, whereas for odd-odd nuclides is given by $M_0 = 2M_p(T_z=0)$. $B(E2)$ values are given in W.u. and matrix elements in $(\text{W.u.})^{1/2}$. An asterisk indicates that these calculations were done with the sdpf.fin interaction. The rest were done using the USD Wildenthal interaction.

| Nucleus | $T_Z$ | $B(E2)_{exp}$ | Reference | $M_{p_{ex}}$ | $M_{0_{ex}}$ | $B(E2)_{th}$ | $M_{p_{th}}$ | $M_{0_{th}}$ |
|---|---|---|---|---|---|---|---|---|
| $^{18}$F | 0 | > 8 | [11] | >2.83 | >5.66 | 4.40 | 2.10 | 4.20 |
| $^{18}$O | 1 | 3.42(9) | [11] | 1.85(2) | | 1.10 | 1.05 | |
| $^{18}$Ne | −1 | 15.85(1.57) | [8] | 3.98(20) | 5.83(20) | 9.90 | 3.15 | 4.20 |
| $^{22}$Na | 0 | 15.33(4.39) | [20] | 3.92(56) | 7.83(1.12) | 15.14 | 3.89 | 7.78 |
| $^{22}$Ne | 1 | 12.65(22) | [20] | 3.56(05) | | 13.24 | 3.64 | |
| $^{22}$Mg | −1 | 24.49(9.33) | [20] | 4.95(94) | 8.51(94) | 17.18 | 4.14 | 7.78 |
| $^{26}$Al | 0 | 12.61(2.41) | [20] | 3.55(34) | 7.10(68) | 12.75 | 3.57 | 7.14 |
| $^{26}$Mg | 1 | 13.42(36) | [20] | 3.66(05) | | 15.26 | 3.91 | |
| $^{26}$Si | −1 | 14.69(1.44) | [7] | 3.83(19) | 7.49(20) | 10.46 | 3.23 | 7.14 |
| $^{30}$P | 0 | 9.02(62) | [20] | 3.00(10) | 6.00(20) | 10.04 | 3.17 | 6.34 |
| $^{30}$Si | 1 | 7.38(36) | [20] | 2.72(07) | | 8.94 | 2.99 | |
| $^{30}$S | −1 | 11.06(95) | [20] | 3.33(14) | 6.05(16) | 11.18 | 3.34 | 6.33 |
| $^{34}$Cl | 0 | 8.38(94) | [20] | 2.90(16) | 5.79(33) | 6.86 | 2.62 | 5.24 |
| $^{34}$S | 1 | 6.10(18) | [20] | 2.47(04) | | 6.16 | 2.48 | |
| $^{34}$Ar | −1 | 6.76(85) | [20] | 2.60(16) | 5.07(16) | 7.59 | 2.76 | 5.24 |
| $^{38}$K | 0 | 1.48(60) | [20] | 1.22(25) | 2.43(50) | 1.71 | 1.31 | 2.61 |
| | | | | | | 1.91* | 1.38* | 2.76* |
| $^{38}$Ar | 1 | 3.32(14) | [20] | 1.82(04) | | 3.84 | 1.96 | |
| | | | | | | 3.28* | 1.81* | |
| $^{38}$Ca | −1 | 2.52(56) | [5] | 1.59(17) | 3.41(18) | 0.43 | 0.65 | 2.61 |
| | | | | | | 0.91* | 0.95* | 2.76* |
| $^{42}$Sc | 0 | 11.78(3.37) | [9] | 3.43(44) | 6.86(88) | 11.65* | 3.41* | 6.83* |
| $^{42}$Ca | 1 | 9.68(26) | [21] | 3.11(04) | | 9.75* | 3.12* | |
| $^{42}$Ti | −1 | 16.25(4.1) | [21] | 4.03(48) | 7.14(48) | 13.73* | 3.71* | 6.83* |

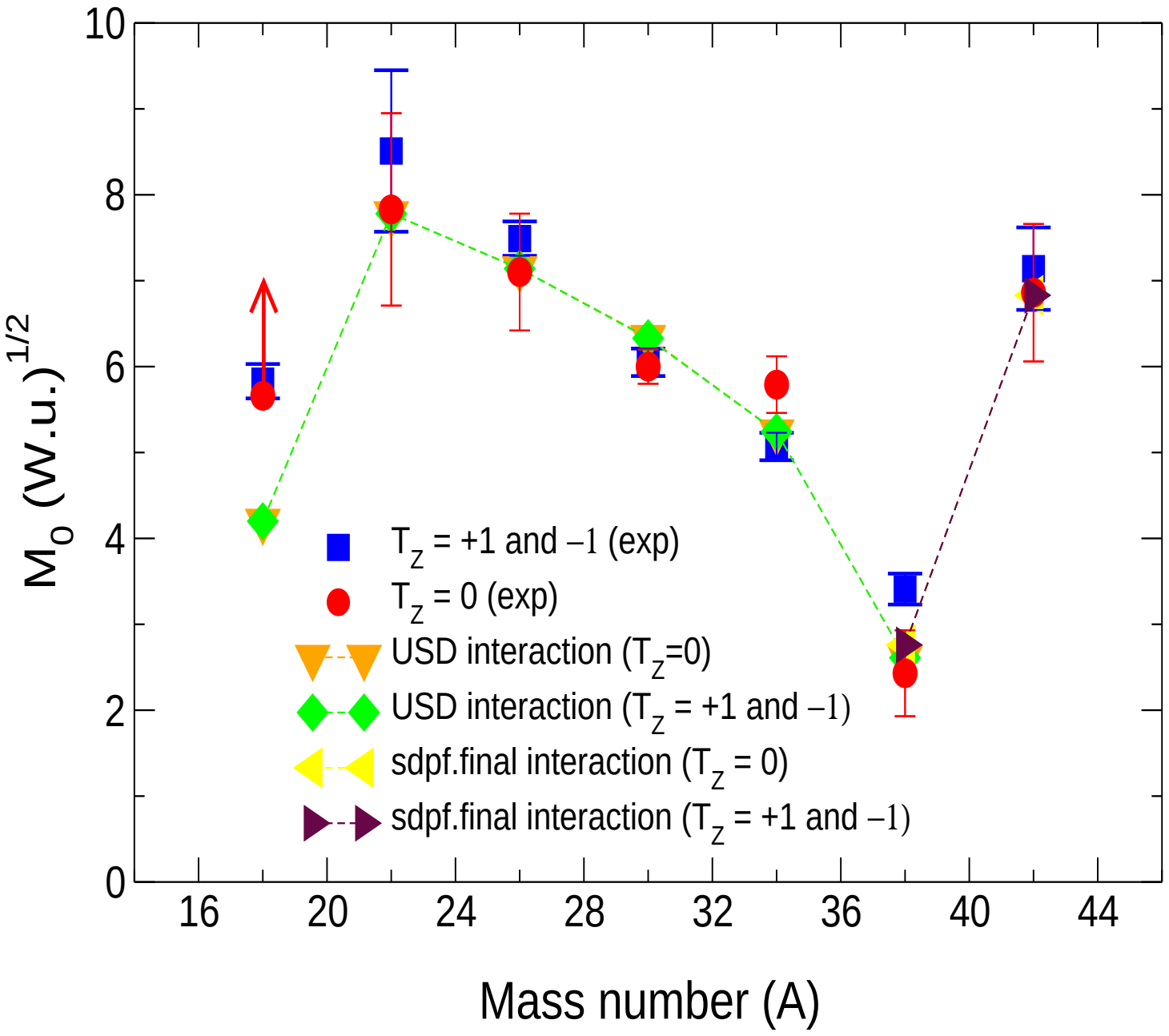

Fig. 2. Experimental and calculated isoscalar multipole matrix elements as a function of mass number for $2_1^+$ $(T = 1) \to 0_1^+$ $(T = 1)$ transitions in the $A = 4n + 2$ isobars ranging from $A = 18$ to $A = 42$. The experimental results are taken from Refs. [5,7–9,11,20,21]

isoscalar matrix elements, $M_0$, for the whole chain of isobaric triplets are presented in Fig. 2. The assumption of isospin symmetry in the nucleon-nucleon interaction leads automatically to the equality of $M_{0_{th}}$ for calculations at each $A$-value (see last column of Table 1). The results are in good agreement with the experimental data, and again, discrepancies arise in the $A = 18$ isobars. This is an interesting result, since these isobars are only few nucleons away from the $^{16}$O core. Similar results were found by Brown *et al.* [12] for $A = 18$ by using shell model calculations with the Chung-Wildenthal interaction. We have used the latter for comparison with the $USD$ interaction, obtaining similar results with both. In fact, indications for broken core in the $A = 18$ system were previously suggested [22,23] and attributed to weakly coupled particle-hole excitations of the $^{16}$O core. Consequently, we have also performed further calculations taking a $2p + 2n$ core, extending the valence space to the $p_{3/2}$, $p_{1/2}$, $d_{5/2}$, $s_{1/2}$ and $d_{3/2}$ orbits, and using the *psd.int* interaction [24]. However, the $B(E2)$ strength predicted for the $2^+_{1\,(T=1)} \to 0^+_{1\,(T=1)}$ transition in $^{18}$F is much lower than the experimentally determined one. Several efforts are being developed with the aim of improving the theoretical predictions, such as large basis *ab initio* calculations [25], model-independent low momentum nucleon-nucleon interactions as the $V_{low\,k}$ [26], or interactions accounting for three-body forces [27].

Discrepancies between experimental and calculated $M_0$ values in the $A = 38$ triplets arise, in this case, at the mirror pair. An idea that might explain these deviations has recently been addressed by Ekman *et al.*, and involves isospin-broken interactions in nuclei. In a study of the neighboring $^{35}$Ar and $^{35}$Cl mirror pair, mirror energy differences and discrepancies in the decay properties of these nuclides were explained in terms of the electromagnetic spin-orbit coupling [28]. Another possibility is suggested from our shell model calculations, which present similar discrepancies at the extremes of the *sd* shell. These discrepancies seem to be closely related to each other since for the $A = 18$ isobaric triplets there are only two particles in the sub-shell above the shell closure, whereas in $A = 38$ triplets there are two holes below the shell closure. This suggests that the low and nearly full occupancy in these orbits at the extremes of the *sd* shell play an important role in the increased collectivity of the experimental $B(E2)$ strengths, through core polarization effects not accomodated by effective charges designed for nuclei throughout the sd shell. These effects might lead to a different $T_Z$ dependence for $A = 18$ and $A = 38$ triplets with respect to the other isobaric triplets within the shell. More reliable experimental data are needed in the $A = 18$ and $A = 38$ isobaric triplets in order to take further conclusions.

## 3 Conclusions

The two notable results of the present calculations is first to see that the progressive decrease of the $M_0$ values through the *sd* shell up to $A = 38$ is well reproduced by the shell model calculations. Secondly, the sharp increase of $M_0$ as the $Z = N = 20$ point is crossed is probably well represented through including core contributions from the *sd* shell into the *fp* valence space, just as those same core excitation contributions solved the earlier problem of $g$ factor issues for the light *Ca* isotopes. Core excitations and the 'weakening' of the N=Z=20 shell boundaries are supported from our calculations in the $A = 38$ system opening the *fp* shell and using the *sdpf.fin* interaction. Despite small deviations for $A = 18$ and $A = 38$, recent experimental data oriented to study $T = 1$ isobaric triplets are well reproduced by the predicted $2^+_{1\,(T=1)} \rightarrow 0^+_{1\,(T=1)}$ transition energies, reduced transition probabilities and isoscalar matrix elements for mirror pairs, $M_0 = M_p(T_Z = -1) + M_p(T_Z = +1)$, and T$_Z$=0 nuclides, $M_0 = 2M_p(T_Z = 0)$. The good agreement between experimental and calculated $M_0$ values provide and explanation of the systematic trend noted in the beginning of this paper, which was the motivation for this work. The discrepancies in the $A = 18$ and $A = 38$ triplets are discussed qualitatively in terms of the occupancy of the orbits immediately above and below shell closures, which might cause a different $T_Z$ dependence. Both experimental and theoretical work on the $A = 18$ and $A = 38$ triplets are necessary to fully understand isospin symmetry at the extremes of the *sd* shell. Finally, future development of radioactive ion beam facilities would allow the study of $A = 4n + 2$ isobaric triplets above $A = 42$. Yamada *et al.* have just carried out the first measurement of reduced transition probabilities for the $2^+_{1\,(T=1)} \rightarrow 0^+_{1\,(T=1)}$ transition in $^{46}$Cr, by intermediate-energy Coulomb excitation in inverse kinematics [29]. This new information could lead to detailed insights of systematics effects and a better understanding of isobaric triplets further along the $N \sim Z$ line.

## 4 Acknowledgements

The authors would like to thank M. T. McEllistrem for his guidance and revision of the current paper, S. W. Yates and F. M. Prados-Estévez for useful discussions, and F. Nowacki for code support. This work was partially supported by the U.S. National Science Foundation under Grant No. PHY-0354656.